\documentclass[a4paper,11pt]{article}
\pdfoutput=1
\usepackage{jheppub}
\usepackage[T1]{fontenc}
\usepackage{amsfonts,amssymb,epsfig,verbatim,mathbbol,amstext,amsmath,psfrag}
\usepackage{graphicx}
\usepackage{t1enc}
\usepackage{subfigure}
\usepackage{dsfont}
\usepackage{slashed}

\usepackage{colordvi}
\usepackage{xcolor}
\usepackage{color}

\usepackage{tikz}
\usetikzlibrary{decorations.pathmorphing,decorations.markings}

\newcommand{\cref}[1]{(\ref{#1})}
\newcommand{\dd}{\textmd{d}}
\newcommand{\be}{\begin{equation}}
\newcommand{\ee}{\end{equation}}
\newcommand{\Tr}{\textmd{Tr}}
\newcommand{\Z}{\mathcal{Z}}

\title{On electric fields in hot QCD: perturbation theory}

\author[a]{G.~Endr\H{o}di, G.~Mark\'o}

\affiliation[a]{Fakult\"at f\"ur Physik, Universit\"at Bielefeld,
D-33615 Bielefeld, Germany}

\emailAdd{endrodi@physik.uni-bielefeld.de}
\emailAdd{gmarko@physik.uni-bielefeld.de}

\abstract{
We investigate the response of a hot gas of quarks to external electric fields via leading-order perturbation theory. In particular, we discuss how 
equilibrium is maintained in the presence of the electric field and 
calculate the electric susceptibility, providing 
its high-temperature expansion for arbitrary quark mass.
Furthermore, we point out that there is a mismatch between this, direct determination of the susceptibility at zero field and the weak-field expansion of the effective action at nonzero electric fields, as obtained using
Schwinger's exact propagator. We discuss the origin of this mismatch and 
elaborate on the generalization of our results to full QCD 
in electric fields.
}


\begin{document} 
\maketitle
\flushbottom

\section{Introduction\label{sec:intro}}

The quantum vacuum is not empty space but a dynamical state containing
virtual particles that can be viewed as a medium. This medium 
can be most efficiently probed by background electromagnetic fields, both 
for the case of the strong interactions -- virtual quarks and gluons described 
by Quantum Chromodynamics (QCD), 
as well as for electromagnetism -- virtual electrons and photons described 
by Quantum Electrodynamics (QED).

For strongly interacting systems, such background electromagnetic fields 
are known to play a relevant role 
for a range of physical settings including magnetized neutron stars, 
the evolution of the early Universe and off-central heavy-ion collisions.
In the latter case, the phenomenological impact of magnetic fields $B$ has been 
recognized primarily
in the context of the chiral magnetic effect~\cite{Fukushima:2008xe}
and collective motion.
These collisions feature magnetic fields that by far
exceed the QCD scale~\cite{Skokov:2009qp,Huang:2015oca},
even though their life-time in the plasma 
is the subject of active debate~\cite{Tuchin:2013apa,McLerran:2013hla,Huang:2015oca}. 
Recently, it was recognized that heavy-ion collisions 
might also generate strong electric fields $E$
on an event-by-event basis, 
with components comparable to the magnetic ones in magnitude~\cite{Deng:2014uja,Voronyuk:2014rna}.
This has been in the focus of recent interest also due to the isobaric experimental 
program at RHIC~\cite{Voloshin:2010ut}. 

In turn, the response of the QED vacuum 
to intense background fields is by now routinely explored 
by high-intensity lasers~\cite{Gies:2008wv,DiPiazza:2011tq}.
These new-generation laser facilities (e.g.\ Ref.~\cite{Yoon:2021ony}) 
operate at unprecedented intensities, probing the structure of the quantum 
vacuum and aiming at recreating astrophysical environments in the laboratory~\cite{Marklund:2006my}. Current efforts also include 
collision experiments involving high-energy electron beams and laser pulses, 
as in the upcoming LUXE experiment~\cite{Abramowicz:2021zja}.
For a recent summary, see e.g.\ Ref.~\cite{Fedotov:2022ely}.

Much of our knowledge about the impact of magnetic fields on QCD matter comes from first-principles
lattice simulations regarding the phase diagram~\cite{DElia:2010abb,Bali:2011qj,Bali:2012zg,Endrodi:2015oba,DElia:2021yvk} and the equation of state~\cite{Bonati:2013vba,Bali:2014kia,Bali:2020bcn}, as well as from a wide range 
of different model and effective theory approaches (for a recent review, see Ref.~\cite{Andersen:2014xxa}). In contrast, there are only a few results about the role of electric fields, mainly due to the fact that the QCD action at $E\neq0$ becomes complex, hindering standard simulation algorithms~\cite{Engelhardt:2007ub,Detmold:2009dx,Yamamoto:2012bd,Yang:2022zob}. Owing to asymptotic freedom in QCD, ordinary perturbation theory is expected to work as a high-temperature alternative of the lattice approach. Nevertheless, convergence problems restrict the region of applicability to very high temperatures \cite{Arnold:1994eb,Braaten:1995ju,Kajantie:2002wa}, unless mended by resumming the perturbative series using, e.g., hard thermal loop techniques \cite{Haque:2014rua}, bridging the gap towards the lattice approach~\cite{Borsanyi:2010cj,HotQCD:2014kol}. In stark contrast, the effect of weak magnetic fields is found to be captured very well already at the one-loop 
order of ordinary perturbation theory
down to $T\approx 300 \textmd{ MeV}$~\cite{Bali:2014kia,Bali:2020bcn}. 
A perturbative treatment at nonzero electric fields is therefore 
worth pursuing and in this paper we will focus on the one-loop order.
In turn, the analogous perturbative calculation is applicable for hot QED as well.

There are two main approaches to study this problem. First, one may use the 
exact fermion propagator at nonzero $E$ due to Schwinger~\cite{Schwinger:1951nm} and construct the free energy density $f(E)$, followed by a weak-field expansion to find the 
leading coefficient, the electric susceptibility $\xi$. This method has 
been employed both with real-time propagators~\cite{Loewe:1991mn,Elmfors:1994fw,Elmfors:1998ee} as well as within the 
imaginary time formalism~\cite{Gies:1998vt,Gies:1999xn} at arbitrary temperatures. Second, one may consider the 
expansion of $f$ in the presence of arbitrary external photon fields to express $\xi$
via the photon vacuum polarization diagram directly at $E=0$. This 
approach goes back to Weldon~\cite{Weldon:1982aq}, who calculated the leading $\mathcal{O}(T^2)$ temperature-dependence for massless fermions.
The above two ways of treating background fields in hot plasmas have 
both been instrumental for a whole set of subsequent calculations. 
On the one hand, Schwinger's approach has been used for constructing Euler-Heisenberg-type effective 
actions~\cite{Dunne:2004nc} in the context of laser physics, for QED thermodynamics as well 
as QCD effective models~\cite{Miransky:2015ava}. On the other hand, Weldon's approach, generalized to color interactions, has formed the basis for hard thermal loop 
perturbation theory~\cite{Braaten:1991gm,Blaizot:2001nr} and was employed 
for various physical applications within quark-gluon plasma physics, see e.g.\ Ref.~\cite{Arnold:2003zc}.

In this paper we follow Weldon's method and calculate the 
electric susceptibility for arbitrary fermion masses and present its high-temperature expansion for the first time within this approach. We discuss the importance and effect of the equilibrium 
charge distribution that forms under the impact of 
the electric field. Furthermore, we point out that 
the so defined susceptibility disagrees with the one obtained via Schwinger's approach, revealing the 
significance of an infrared regularization for this observable and the 
non-commutativity of the involved limits (zero electric field and infinite volume). Finally, we discuss how our results 
can be generalized to full QCD.

\section{Susceptibilities, renormalization and equilibrium{\label{sec:defs}}}

In the presence of weak background electromagnetic fields,
QCD matter in thermal equilibrium 
may be treated as a linear medium. The total free energy density, considered 
as a function of the thermodynamic variables $E$ and $B$, reads
$f^{\rm tot}=B^2/(2\mu)-\epsilon E^2/2$, where $\mu$ is
the magnetic permeability and $\epsilon$ the electric permittivity ~\cite{landau1995electrodynamics}. 
These parameters enter the constitutive relations $B=\mu H$ and 
$D=\epsilon E$. In the following we refer to 
$B$ and $E$ simply as magnetic and electric fields.

The deviation from the vacuum values ($\epsilon=\mu=1$ in natural units)
can be measured by the electric and magnetic susceptibilities as
$\epsilon=1+e^2\xi$ and $\mu=(1-e^2\chi)^{-1}$. The factor $e^2$ -- the square of the elementary charge -- is included here for later convenience.
Below we exclude the energy $f^\gamma=(B^2-E^2)/2$ of the fields themselves
and merely consider the matter contribution $f=f^{\rm tot}-f^{\gamma}$. 
In terms of the partition function $\Z$ of the system, it is given by 
$f=-T \log\Z/V$, where $T$ is the temperature and $V$ the spatial volume.
The susceptibilities are then defined as
\be
\xi_b=-\left.\frac{\dd^2 f}{\dd (eE)^2}\right|_{E=0}\,, \quad\quad
\chi_b=-\left.\frac{\dd^2 f}{\dd (eB)^2}\right|_{B=0}\,.
\label{eq:chidef}
\ee
Note that at $T=0$, Lorentz invariance ensures that $\chi_b=-\xi_b$. 
In Eq.~\eqref{eq:chidef} the subscript $b$ indicates that both susceptibilities contain ultraviolet divergent terms that must be 
subtracted via additive renormalization. These divergences originate 
from the multiplicative divergence in the electric charge $e$~\cite{Schwinger:1951nm,Dunne:2004nc},
\be
B^2=Z_e B_r^2, \qquad E^2=Z_e E_r^2, \qquad
e^2=Z_e^{-1} e_r^2\,,
\label{eq:Ze1}
\ee
with the renormalization constant which up to $\mathcal{O}(e_r^2)$ reads
\be
Z_e=1+\beta_1 e_r^2 \log \frac{\sigma^2}{\Lambda^2}\,,\qquad \beta_1=\frac{1}{12\pi^2}\,,
\label{eq:Ze2}
\ee
where $\Lambda$ is an UV cutoff and $\sigma$ the renormalization scale.
Note that the QED Ward identity ensures that $e_rB_r=eB$ and $e_rE_r=eE$ are renormalization group invariant combinations. 

To see how this is 
related to additive divergences in the bare susceptibilities, consider 
the total free energy density, which is a finite, physical quantity, 
\be
f^{\rm tot}=\frac{B^2-E^2}{2} - \xi_b\frac{(eE)^2}{2} - \chi_b \frac{(eB)^2}{2} + \mathcal{O}(E^4, B^4, E^2B^2)\,.
\ee
For this to be finite, Eqs.~\eqref{eq:Ze1} and~\eqref{eq:Ze2} imply that
\be
\xi_b= \beta_1 \log \frac{\sigma^2}{\Lambda^2} + \textmd{ finite}\,, \qquad
\chi_b= -\beta_1 \log \frac{\sigma^2}{\Lambda^2} + \textmd{ finite}\,.
\ee
Being 
temperature-independent, the additive divergence cancels in
\be
\xi = \xi_b(T)-\xi_b(T=0)\,, \qquad
\chi=\chi_b(T)-\chi_b(T=0)\,,
\label{eq:xirdef}
\ee
which sets $\xi=\chi=0$ at zero temperature after renormalization, consistent with the requirement 
that the electric permittivity $\epsilon$ and magnetic permeability $\mu$ of the vacuum both be unity.
For a recent summary on the role of electric charge renormalization for
susceptibilities and on the choice of the renormalization 
scale $\sigma$, see Ref.~\cite{Bali:2020bcn}. 

\begin{figure}[t]
 \centering
 \includegraphics[width=8cm]{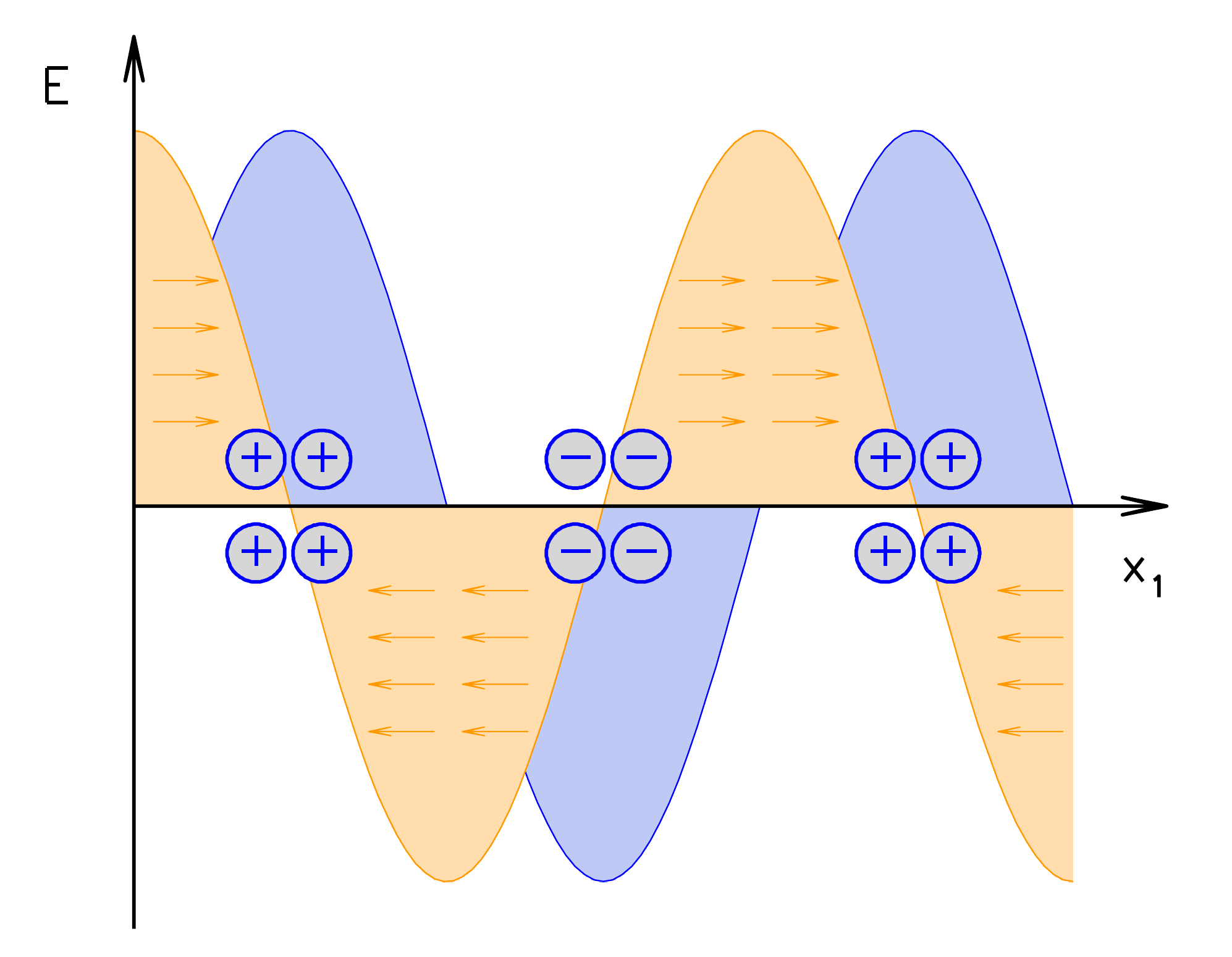}
 \caption{\label{fig:demo1}Illustration of the equilibrium situation in the presence of a weak
 electric field $E \parallel x_1$ modulated in the $x_1$ direction 
 (dark yellow area and arrows). 
 Positive (negative) electric charges accumulate at points with $E=0$ pointed 
 towards (away from) the electric field, giving rise to the charge density profile 
 indicated by the blue region.
}
\end{figure}

Above we assumed $E$ and $B$ to be homogeneous. While for the magnetic field this assumption is consistent with equilibrium, a constant 
electric field accelerates charged particles indefinitely and inevitably leads the system out of equilibrium. 
In other words, homogeneous electric fields necessarily involve some type of infrared regularization, for example a finite volume with boundaries, where charges can accumulate. Alternatively, one can consider spatially oscillating electric fields and their infinite wavelength limit. For such a (time-independent) electric field, the system will find the equilibrium configuration, where electric and diffusion forces cancel each other and no currents flow\footnote{This point is also discussed in Ref.~\cite{Luttinger:1964zz}. Note also that our electromagnetic fields are classical background fields, therefore the charges only interact with $E$ or $B$ but not with each other.}, illustrated in Fig.~\ref{fig:demo1}. 
In this setup, the impact of the electric field is therefore to displace electric charges and to polarize the medium. Rotational invariance and the homogeneity of the $E=0$ medium ensures that for weak fields
the polarization oscillates with the same wavelength as the electric field does. We note that this interpretation also holds for time-dependent fields, as long as the corresponding time-like frequency approaches zero before the inverse wavelength does.\footnote{In contrast, the opposite case of an electric field with infinite wavelength but nonzero time-like frequency leads to a quasi-equilibrium, where currents are flowing back and forth in an oscillatory fashion. This setting is described by a different type of linear response, namely the electric conductivity of the medium.}

Without loss of generality, the electric field is assumed to point in the $x_1$ direction and from now on, $E$ denotes this non-vanishing component of the field unless otherwise noted. The electric field couples to the quark charges $q_f$ via the vector potential $A_\nu$, for which we choose the static gauge $A_0(x_1)$ and $\mathbf{A}=0$. A constant electric field is then realized by $A_0=Ex_1$, while an example oscillating field with a well defined equilibrium is $E(x_1)=E\cdot \cos(k_1x_1)$, generated by $A_0(x_1)=E\cdot \sin(k_1x_1)/k_1$ with spatial momentum $k_1$, illustrated in Fig~\ref{fig:demo1}. For completeness, we will also include results for the magnetic susceptibility $\chi$. 
For the latter, we choose a gauge potential 
$A_2(x_1)$, so that $B$ points in the $x_3$ direction, however we always consider electric and magnetic fields separately. In the following we calculate the susceptibilities for a single fermion with charge $e$ -- the result for QCD is obtained via multiplication by the number $N_c=3$ of colors and by $\sum_f(q_f/e)^2$, where $f$ runs over the quark flavors.

\section{Calculation of the electric susceptibility}
The definition~\eqref{eq:chidef} of the electric susceptibility can be evaluated in two different ways, as already pointed out in Sec.~\ref{sec:intro}. We will first obtain $\xi$ by evaluating the derivatives of the free energy directly at $E=0$ using the expansion of the partition function in terms of arbitrary but weak fields \cite{Weldon:1982aq}, and then we will compare this result to Refs.~\cite{Loewe:1991mn,Elmfors:1994fw,Elmfors:1998ee}, where the free energy is calculated at nonzero but homogeneous $E$ and the susceptibility is read off from its weak field expansion. Throughout we are using one-loop perturbation theory. For definiteness, the metric tensor is $g_{\mu\nu}=\textmd{diag}(1,-1,-1,-1)$ and four-vectors are denoted as for example $K=(k_0,\mathbf{k})$ with $k=|\mathbf{k}|$ and $K^2=k_0^2-k^2$.

\subsection{Expansion at arbitrary small fields}

As proposed in Sec.~\ref{sec:defs}, in the presence of the electric field the equilibrium situation involves a fixed charge density profile $n(x_1)$ that varies in the $x_1$ direction in the system (of size $L$). This follows from the equilibrium requirements of a statistical system in a background field, see e.g. \cite{landau2013statistical}. In this setup the temperature and the total chemical potential are constant, however the pressure and the particle density are inhomogeneous.
We will first discuss the implications of such an equilibrium using a homogeneous background field regularized by the finite system size, rather than by oscillating with a well defined wavelength. The generalization for the oscillating fields will follow, however we found it more didactic to present this matter in two steps. 

\subsubsection{Impact of the charge distribution}
\label{sec:chargedistr}

For weak fields, the gradient of $n(x_1)$ is small so that one can imagine the system as a collection of smaller (but still large) subsystems with each subsystem having a given constant density. This is then described by a {\it canonical} thermodynamic potential $f$ that depends on the local density and is related to the grand canonical potential $\Omega$ via an $x_1$-dependent Legendre transform
\be
\label{eq:localLegendre}
-f(E,n(x_1)) = \frac{1}{L}\int d x_1 \left[-\Omega(E,\mu)-\mu n(x_1)\right]_{\mu=\bar\mu(x_1)}\,, \qquad
n(x_1)=-\left.\frac{\partial \Omega(E,\mu)}{\partial \mu}\right|_{\mu=\bar\mu(x_1)}\,,
\ee
trading the dependence on a local chemical potential $\bar\mu(x_1)$ 
for the dependence on $n(x_1)$. The fixed density is such that the resulting diffusion force
balances the electric field: $\partial\bar\mu(x_1)/\partial x_1=-eE$, thus
\be
\label{eq:equiMu1}
\bar\mu(x_1) = -eA_0(x_1)\,.
\ee
In other words, the sum of the local chemical potential due to the accumulation of charges and the external potential is constant~\cite{landau2013statistical}.
A constant shift in $\bar\mu(x_1)$ would merely change the total charge in the system -- we set this to zero assuming a neutral medium.

Thus, the canonical potential $f$, relevant for the equilibrium situation, depends
on $E$ in two ways: explicitly as well as 
implicitly via the charge distribution. This 
implicit dependence reads
\be
\frac{\partial n(x_1)}{\partial E} = -\left.\frac{\partial}{\partial E}\frac{\partial \Omega(E,\mu)}{\partial \mu}\right|_{\mu=\bar\mu(x_1)}=-\left.\frac{\partial ^2 \Omega(E,\mu)}{\partial E\partial \mu}\right|_{\mu=\bar\mu(x_1)}-\left.\frac{\partial^2 \Omega(E,\mu)}{\partial \mu^2}\right|_{\mu=\bar\mu(x_1)}\cdot\frac{\partial \bar\mu(x_1)}{\partial E}\,.
\ee
Altogether, the susceptibility~\eqref{eq:chidef} as the second total derivative with respect to $E$, becomes
\be
\label{eq:xi_equi}
\xi_b = -\left.\frac{\dd^2 f}{\dd (eE)^2}\right|_{E=0}=-\frac{1}{L}\int d x_1\left[\frac{\partial^2 \Omega}{\partial (eE)^2}-\frac{\partial^2 \Omega}{\partial\mu^2}\cdot\left(\frac{\partial\bar\mu(x_1)}{\partial (eE)}\right)^2\right]_{E=\mu=0}\,,
\ee
where we used that $\Omega$ is an even function both of $\mu$ and of $E$.
Notice that the first term is the curvature of $\Omega$with respect to the electric field at $\mu=0$, i.e.\ without the equilibrated charge distribution. In turn, the second term is related to the free energy of the displaced charges themselves. For homogeneous fields $\bar\mu(x_1)=-eE x_1$ and this term is quadratically divergent in the system size, $\propto L^2$. We will find that the first term contains the same divergence with an opposite sign so that $\xi_b$ is altogether infrared finite.

\subsubsection{Susceptibilities for oscillatory fields}

Now we generalize \eqref{eq:xi_equi} for the case of oscillating fields. In the spirit of the end of Sec.~\ref{sec:defs}, we consider the general linear response of a translationally and rotationally symmetric system
\be
\label{eq:xiMomFreeEn}
f = - \frac{e^2}{2}\frac{T}{V}\int \frac{d^4 K}{(2\pi)^4} \,E(K) \,\xi_b(K)\, E(-K)+{\cal O}(E^4)\,,
\ee
thus redefining the electric susceptibility as the limit
\be
\label{eq:xiMomDef}
\xi_b = \lim_{k\to0}\lim_{k_0\to0}\xi_b(K)\,,
\ee
where the time-like frequency $k_0$ approaches zero first, as discussed in Sec.~\ref{sec:defs}.

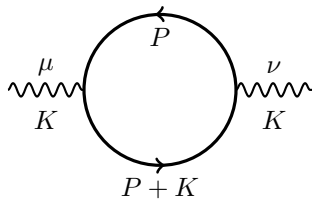
\begin{figure}[ht]
 \centering
 \begin{tikzpicture}[ thick,font=\large ]
 \draw[black, very thick,decoration={markings, mark=at position 0.26 with {\arrow{>}}, mark=at position 0.76 with {\arrow{>}}}, postaction={decorate}] (0,0) circle (1cm);
 \draw[black,decorate, decoration={snake,segment length=6pt, post length=0pt}] (-2,0) -- (-1,0);
 \node at (-1.5,0.3) {\small $\mu$};
 \draw[black,decorate, decoration={snake,segment length=6pt, post length=0pt}] (1,0) -- (2,0);
 \node at (1.5,0.3) {\small $\nu$};
 \node at (-1.5,-.4) {\small $K$};
 \node at (1.5,-.4) {\small $K$};
 \node at (0,.7) {\small $P$};
 \node at (0,-1.3) {\small $P+K$};
 \end{tikzpicture}
 \caption{\label{fig:polTens}Vacuum polarization diagram. Since the background electric field is classical, this is the only diagram contributing to the susceptibility. The solid lines denote the finite temperature fermion propagators.}
\end{figure}

To find $\xi_b$ in this setup, we need $\Omega$ for an arbitrary background gauge field $A^\alpha$ in order to express
the derivatives appearing in Eq.~\eqref{eq:xi_equi}.
To this end, we write down the leading order expansion of $\Omega$ for general background gauge fields~\cite{Weldon:1982aq}, 
\be
\label{eq:omegaDeriv}
\Omega = -\frac{1}{2}\int \frac{d^4K}{(2\pi)^4} \,\widetilde{A}^\alpha(K) \,\Pi_{\alpha\beta}(K) \,\widetilde{A}^\beta(-K) + \mathcal{O}(\widetilde{A}^4)\,,
\ee
where $\widetilde{A}^\alpha(K)$ is the momentum-space gauge field and $\Pi_{\mu\nu}(K)$ the photon vacuum polarization tensor, see Fig.~\ref{fig:polTens}. 
Based on Eq.~\eqref{eq:omegaDeriv}, we can evaluate the partial derivatives in Eq.~\eqref{eq:xi_equi} by choosing specific gauge fields representing either the oscillating electric field $E(x_1)=E\cdot\cos(k_1x_1)$ or the chemical potential. Explicitly, for the $\partial^2 \Omega/\partial (eE)^2$ term appearing in \eqref{eq:xi_equi}, we have to consider $\widetilde{A}^\alpha(K) =  \delta^{\alpha 0} E(K)/(ik_1)$. In turn, for $\partial^2 \Omega/\partial \mu^2$ we need $\widetilde{A}^\alpha(K) = \delta^{\alpha 0}\mu\, \delta^{(4)}(K)$, and finally $(\partial\bar\mu(x_1)/\partial (eE))^2$ becomes $\frac{\delta A_0(x_1)}{\delta E(K)}\frac{\delta A_0(x_1)}{\delta E(-K)}=1/k_1^2$. Combining all this we find
\be
\xi_b(K) = \frac{1}{L}\int dx_1\,\frac{1}{e^2} \left[\frac{\Pi_{00}(k_0,k)}{k_1^2}-\frac{\Pi_{00}(k_0=0,k=0)}{k_1^2}\right]\,,
\ee
where in the second term it is understood that $k_0$ approaches zero before $k$ does.
Since nothing depends on $x_1$ anymore, the averaging is trivial. Finally, we subtract the zero temperature part to renormalize as prescribed in \eqref{eq:xirdef},
\be
\label{eq:xiDefWPi}
\xi = \frac{1}{e^2}\lim_{k\to0}\lim_{k_0\to0} \left[\frac{\Pi_{00}^{T\neq 0}(k_0,k)-\Pi_{00}^{T\neq 0}(0,0)}{k^2}\right]
= \frac{1}{2e^2}\lim_{k\to0}\lim_{k_0\to0}\frac{\partial^2 \Pi_{00}^{T\neq0}(k_0,k)}{\partial k^2}\,,
\ee
where $\Pi_{\mu\nu}^{T\neq0}=\Pi_{\mu\nu}^{T}-\Pi_{\mu\nu}^{T=0}$ is the thermal part of the vacuum polarization diagram. Here we replaced $k_1^2$ by $k^2$, restoring the rotational symmetry of the system, which we explicitly broke by choosing $E$ to point in a particular direction. 
Moreover, we rewrote the $k\to0$ limit of the finite difference as half of the second derivative with respect to $k$. This is legitimate, as $\Pi_{00}^{T\neq0}$ has no linear term in $k$ around $k=0$ due to parity symmetry. The first term in the square brackets of Eq.~\eqref{eq:xiDefWPi} agrees
with Weldon's formula~\cite{Weldon:1982aq}, while 
the second term stems from the local charge distribution, which was not taken into account in Ref.~\cite{Weldon:1982aq}.

The analogous expression for the magnetic susceptibility does not have the second term, since a modified equilibrium charge distribution is not formed for purely magnetic backgrounds. For the gauge $A_2(x_1)$ we chose for the magnetic field, the component $\Pi_{22}$ shows up in the magnetic susceptibility based on \eqref{eq:omegaDeriv}. One can restore rotational symmetry by including the other spatial components (see e.g.~\cite{Weldon:1982aq}), and the magnetic susceptibility becomes
\be
\chi = \frac{1}{e^2}\lim_{k\to0}\lim_{k_0\to0}\frac{\Pi_S^{T\neq0}(k_0,k)}{k^2}
= \frac{1}{2e^2}\lim_{k\to0}\lim_{k_0\to0}\frac{\partial^2 \Pi_S^{T\neq0}(k_0,k)}{\partial k^2}\,.
\label{eq:chiDefWPi}
\ee
where $\Pi_S = \sum_{i=1}^3\Pi_{ii}/2$. Unlike the electric component of the gauge field, the magnetic ones do not acquire thermal masses and therefore $\Pi_S^{T\neq0}(0,k)\sim k^2$ in the low-momentum region. Thus we 
could replace division by $k^2$ by half of the second derivative with respect to $k$, just as above.
Eqs.~\eqref{eq:xiDefWPi} and~\eqref{eq:chiDefWPi} are our final formulas for the susceptibilities. Now we proceed by calculating the vacuum polarization tensor at finite temperature.

\subsubsection{Photon vacuum polarization diagram}

To calculate the thermal part $\Pi_{\mu\nu}^{T\neq 0}(k_0,k)$, we follow and extend the calculation of Weldon \cite{Weldon:1982aq} to arbitrary fermion masses $m$. The general formula is
\be
\label{eq:polTensWSF}
\Pi_{\mu\nu}(k_0,k) = i e^2\int \frac{d^4 P}{(2\pi)^4} {\rm \,Tr\,} \left[\gamma_\mu S_F(P+K)\gamma_\nu S_F(P)\right]\,,
\ee
where the fermion propagators running in the loop are the finite temperature Feynman propagators
\be
S_F (P) =(\slashed{P}+m)\left[\frac{1}{P^2-m^2+i\epsilon}+2\pi i n_F(|p_0|)\delta(P^2-m^2)\right]\,,
\ee
with $n_F(|p_0|) = (e^{\beta |p_0|}+1)^{-1}$ being the Fermi-Dirac distribution. We relegate the details of the calculation to App.~\ref{app:Pi00}; here we only summarize the results. Notice that in the limit $k\to0$ when evaluating $\xi$, the leading contribution $\Pi_{00}^{T\neq 0}(0,0)$, i.e.\ the thermal photon mass, drops out and $\xi_b$ is infrared finite. The exact value of this term we merely use to check against previous results (see \eqref{eq:Pi00_00} and \eqref{eq:Pi00_00_HTE}). Nevertheless, since the thermal mass is non-vanishing, should one neglect the $E$-dependence of the equilibrium charge distribution, the susceptibility would be infrared divergent, as we already mentioned above in Sec.~\ref{sec:chargedistr}. This infrared divergence was already observed by Weldon \cite{Weldon:1982aq} (in the massless case), but its interpretation in terms of the equilibrium charge distribution was not discussed.

The second order term in the expansion of $\Pi_{00}^{T\neq0}(0,k)$ around $k=0$ gives then the susceptibility, which reads (see the details again in App.~\ref{app:Pi00})
\be
\frac{\partial^2 \Re \,\Pi_{00}^{T\neq0} (0,k)}{\partial k^2}\Bigg|_{k=0} = -\frac{e^2}{3\pi^2}\int_m^\infty \frac{d\omega}{\sqrt{\omega^2-m^2}}\left(2n_F(\omega)-\omega\frac{dn_F(\omega)}{d\omega}\right)\,,
\ee
and hence
\be
\label{eq:xiRes}
\xi = -\frac{1}{6\pi^2}\int_m^\infty \frac{d\omega}{\sqrt{\omega^2-m^2}}\left(2n_F(\omega)-\omega\frac{dn_F(\omega)}{d\omega}\right)\,.
\ee
One can expand in terms of $m/T$ to obtain a high-temperature expansion (HTE) expression
\be
\xi = -\frac{1}{12\pi^2}\left(\log\frac{T^2\pi^2}{m^2}-2\gamma_E+1\right)+\frac{31\zeta'(-4)}{192\pi^2}\frac{m^4}{T^4}+{\cal O}\left(\frac{m^6}{T^6}\right)\,,
\label{eq:WeldonxiHTE}
\ee
where $\zeta'(z)$ is the derivative of the Riemann $\zeta$ function. The leading term in~\eqref{eq:WeldonxiHTE} 
agrees with the results of Refs.~\cite{Weldon:1982aq,Carignano:2017ovz} for massless fermions -- in that case $m$ is replaced by the renormalization scale under the logarithm.

For the magnetic case using \eqref{eq:ddkPiS} from App.~\ref{app:Pi00} we find
\be
\label{eq:chiRes}
\chi = \frac{1}{3\pi^2}\int_m^\infty d\omega\,\frac{n_F(\omega)}{\sqrt{\omega^2-m^2}}\,,
\ee
which can be expanded in $m/T$ to obtain
\be
\chi = \frac{1}{12\pi^2}\left(\log\frac{T^2\pi^2}{m^2}-2\gamma_E\right)-\frac{7\zeta'(-2)}{12\pi^2}\frac{m^2}{T^2}+{\cal O}\left(\frac{m^4}{T^4}\right)\,.
\ee

\subsection{Background field approach}
In this subsection we will summarize the results described in detail in Ref.~\cite{Elmfors:1998ee}, with similar calculations carried out in e.g.\ Refs.~\cite{Loewe:1991mn,Gies:1998vt,Gies:1999xn}. The method relies on the calculations of Schwinger \cite{Schwinger:1951nm}, where the exact fermion propagator in a constant electromagnetic background was obtained and used to evaluate the one loop effective potential in the vacuum. The extension of this method to finite temperature is carried out in \cite{Elmfors:1998ee}, providing a separation of the effective potential at nonzero $E$ (and $B$) into a vacuum and a temperature-dependent part. We concentrate here on the latter. The one loop effective action is then obtained as a series
\be
f(E) = \frac{1}{4\pi^2}\int_{-\infty}^\infty d\omega\,\frac{\Theta(\omega^2-m^2)}{\sqrt{\omega^2-m^2}}\sum_{n=0}^\infty e^{2n}f_{2n}(\omega)\,,
\ee
where $f_{2n}(\omega)$ is of the order $2n$ in the electromagnetic fields. Therefore we only need $n=1$ to read off the susceptibility\footnote{Note that this already corresponds to the second {\it total} derivative with respect to $E$.}
\be
f_2(\omega) = (E^2-B^2)\frac{n_F(\omega)}{3}+E^2\frac{\omega}{6}\frac{dn_F(\omega)}{d\omega}\,,
\ee
leading to
\be
\label{eq:xiSch}
\xi = -\frac{1}{6\pi^2}\int_m^\infty \frac{d\omega}{\sqrt{\omega^2-m^2}}\left(2n_F(\omega)+\omega\frac{dn_F(\omega)}{d\omega}\right)
\ee
and
\be
\chi = \frac{1}{3\pi^2}\int_m^\infty d\omega\,\frac{n_F(\omega)}{\sqrt{\omega^2-m^2}}\,.
\ee
The expression for $\chi$ agrees with \eqref{eq:chiRes}, 
demonstrating the equivalence of the expansion approach
and the background field approach for the magnetic case.

In contrast, the same is not true for the electric response:
in $\xi$ the term involving the derivative of the Fermi-distribution comes with the opposite sign as in the expansion approach, \eqref{eq:xiRes}. The formula \eqref{eq:xiSch} leads 
to the high-temperature expansion 
\be
\xi = -\frac{1}{12\pi^2}\left(\log\frac{T^2\pi^2}{m^2}-2\gamma_E-1\right)
+\frac{7\zeta'(-2)}{6\pi^2}\frac{m^2}{T^2}
-\frac{31\zeta'(-4)}{128\pi^2}\frac{m^4}{T^4}+{\cal O}\left(\frac{m^6}{T^6}\right)\,,
\label{eq:SchwingerxiHTE}
\ee
differing from the result~\eqref{eq:WeldonxiHTE} of the expansion approach. Most prominently, the sign of the $\mathcal{O}(m^0)$ term next to the logarithmic term is the opposite here\footnote{
The two methods therefore lead to opposite signs for 
the sum $\chi+\xi$. While it is positive for the background field approach, it is negative for the small field expansion.
As well known~\cite{jackson1975classical}, this sum of susceptibilities enters the speed of light in the medium, $v^2/c^2=1/(\epsilon\mu)=1-e^2(\xi+\chi)$, see also Ref.~\cite{Gies:1999xn}. 
As a light wave with momentum $K$ propagates, the electric field that it involves attempts to polarize the medium. The equilibrium charge distribution that we discussed in this paper is not reached in this case. Thus, one should exclude the term due to the nontrivial charge profile from $\xi$, leading to $\xi=\lim_{k\to0}m_\gamma^2/k^2$ with $m_\gamma^2=\Pi_{00}^{T\neq0}/e^2=T^2/3$, see App.~\ref{app:Pi00}. This turns the sign of $\chi+\xi$ positive. In fact, $\xi$
diverges in this case for $k\to0$, signalling that such 
waves cannot enter the thermal medium due to the nonzero photon mass $m_\gamma$. For a more detailed discussion on light speeds in media, see also Ref.~\cite{Latorre:1994cv}.
}.
Notice that 
the two approaches differ by the order of two distinct limits -- while in the small field expansion approach the infrared regulator (the external momentum) is sent to zero after $E\to0$, here the opposite is done, since no explicit long wavelength regulator had to be introduced.

\section{Numerical comparison}

Next we check the analytical one-loop results within the small field expansion by means of numerical calculation of the free fermion determinant on the lattice.
To this end, we need to perform a Wick rotation to imaginary times. Using the particular gauge choices
$A_4=iE x_1$ and $A_2=Bx_1$, the sum of the electric and magnetic 
susceptibilities, 
Eqs.~\eqref{eq:xiDefWPi} and~\eqref{eq:chiDefWPi}, becomes
\begin{equation}
\xi_b+\chi_b= \frac{1}{2e^2}\frac{\partial^2}{\partial k_1^2} \left[ \Pi_{22}^{T\neq0}(0,k_1) - \Pi_{44}^{T\neq0}(0,k_1)\right]_{k_1=0}\,.
\end{equation}
It is advantageous to work with this sum, because 
it is free of additive divergences, since at strictly zero temperature $\chi_b(T=0)=-\xi_b(T=0)$ due to Lorentz invariance. Therefore the sum of the renormalized susceptibilities can be calculated without explicit $T=0$ subtractions, cf.\ Eq.~\eqref{eq:xirdef}. Going over to coordinate space, we obtain
\begin{equation}
\xi+\chi=\frac{1}{2e^2} \int_{-L/2}^{L/2} d x_1 \, x_1^2 \,G(x_1)\,, \qquad
G(x_1)=\int dx_2 dx_3 d x_4 \, \left[\, \Pi_{44}(X)-\Pi_{22}(X)\right]\,,
\label{eq:chiplusxi}
\end{equation}
involving the mixed-representation current-current correlator combination $G(x_1)$. Here we also highlighted
that our numerical calculations are carried out in a finite volume of linear spatial extent $L$. Placing the origin of the coordinate system in the middle makes the formula~\eqref{eq:chiplusxi} consistent with periodic boundary conditions. Truncating the integral at $\pm L/2$
introduces exponentially small errors, because both components of the coordinate-space vacuum polarizations (i.e.\ current-current correlators) decay 
exponentially with the distance. This point is 
discussed in detail for example in Ref.~\cite{Bali:2020bcn}.

We work with the staggered fermion discretization for 
the Dirac operator on high-temperature lattices $N_s^3\times N_t$ with spacing $a$ at nonzero fermion mass $m$. The spatial size and the temperature are $L=N_sa$ and $T=1/(N_ta)$, respectively. 
We calculate $G(x_1)$ using a few thousand random wall sources~\cite{Bali:2015msa}. Since $\Pi_{44}$ and $\Pi_{22}$ correlate with each other, it is advantageous to take their difference for each random source to cancel some of the statistical noise. 
The continuum limit of the result is defined by keeping $LT$ and $T/m$ fixed and sending $N_s,N_t\to\infty$. In turn, the thermodynamic limit is obtained by fixing $ma$ and $Ta=1/N_t$ and sending only $N_s\to\infty$. 

\begin{figure}
    \centering
    \includegraphics[width=0.465\textwidth]{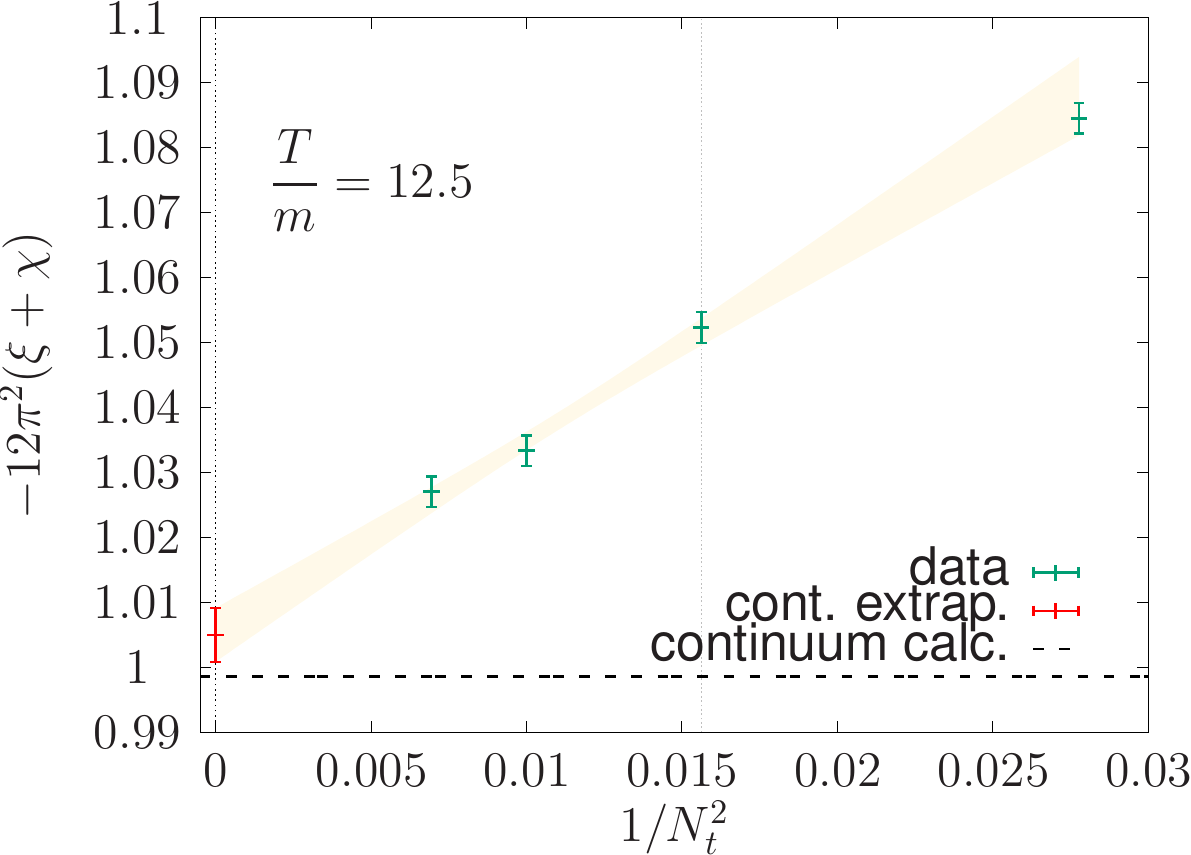}\quad
    \includegraphics[width=0.45\textwidth]{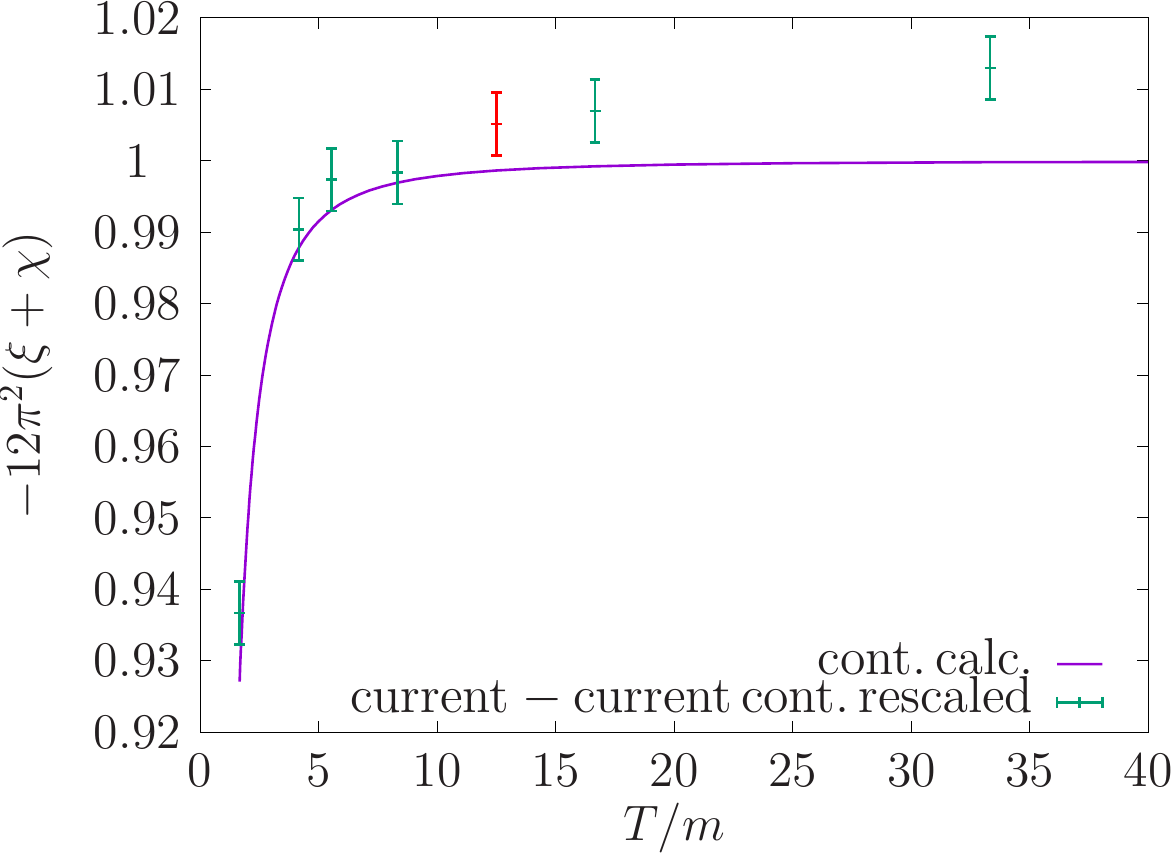}
    \caption{Left: continuum extrapolation of $\xi+\chi$, obtained using Eq.~\protect\eqref{eq:chiplusxi} at $T/m=12.5$ on $LT=4$ lattices (the fit involves only the points up to the grey dotted line), compared to our analytical result from Eqs.~\protect\eqref{eq:xiRes} and~\protect\eqref{eq:chiRes}.
    Right: dependence of the numerical results obtained on $24^3\times6$ lattices (after a rescaling, see the text) on $T/m$, compared to the analytical formula.}
    \label{fig:compare}
\end{figure}

We consider a high temperature with $T/m=12.5$. The continuum extrapolation on $LT=4$ lattices is shown in the left panel of Fig.~\ref{fig:compare}, revealing perfect agreement with the analytical prediction. Note that only the three smallest lattice spacings were used in the extrapolation, indicated by the grey dotted line on the plot. We also compare different spatial sizes and do not observe significant finite volume effects. The ratio of the continuum extrapolation and the $N_t=6$ result is used as an improvement factor. The right panel of Fig.~\ref{fig:compare} shows the temperature-dependence of the $24^3\times6$ data, after a multiplication by this improvement factor. The small deviations from the analytical result are explained by temperature-dependent lattice artefacts. In summary, the numerical data fully confirm our analytical findings.

\begin{figure}
    \centering
    \includegraphics{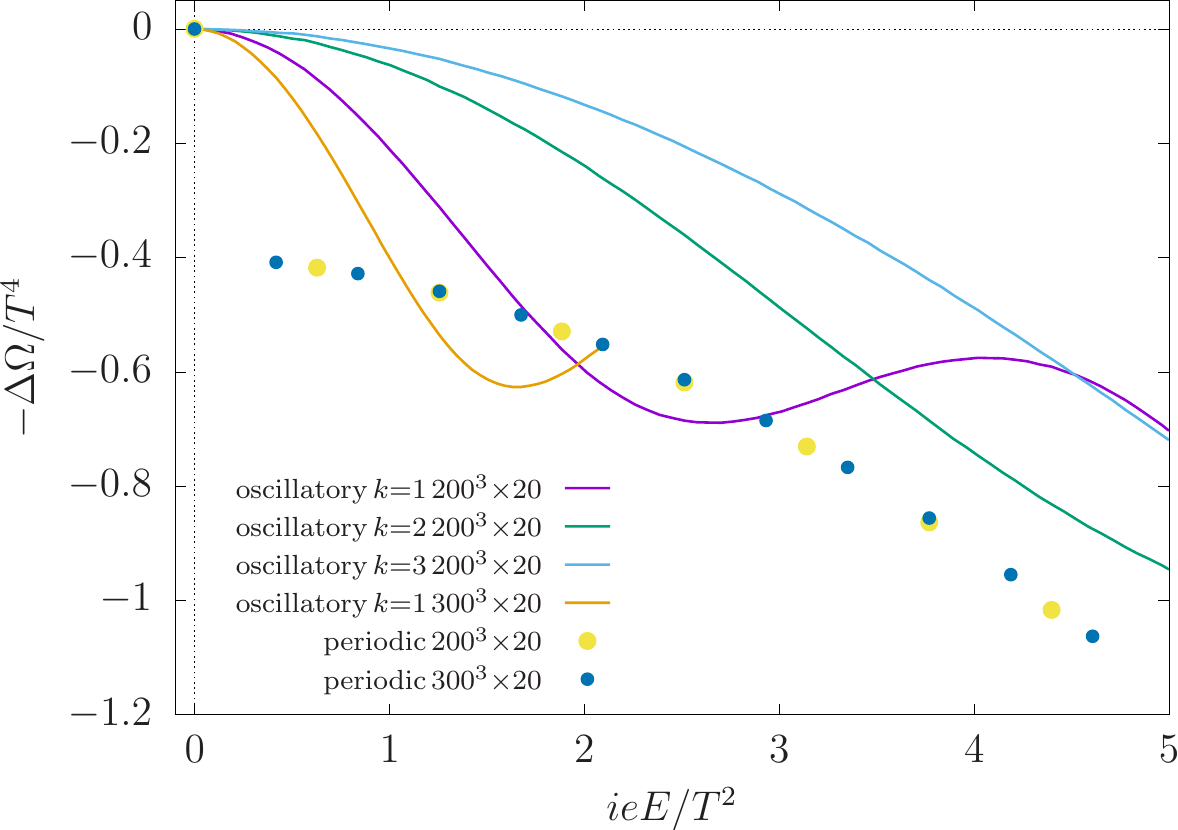}
    \caption{The imaginary electric field dependent part of the grand canonical free energy density as a function of the amplitude of the field strength. The lines represent oscillatory electric fields with different wavelengths, while the dots mark the results for homogeneous fields. Both setups show that towards the thermodynamic limit the grand canonical free energy density develops a discontinuous behaviour at $E=0$.}
    \label{fig:iedep}
\end{figure}
Finally, we comment on lattice simulations at nonzero (imaginary) electric fields and nonzero temperature. As the analytic continuation of the setup with real fields, the equilibrium charge distribution in this case is also inhomogeneous\footnote{See the recent study in Ref.~\cite{Yang:2022zob}, where oscillatory Polyakov loops were observed, and also the similar results in Ref.~\cite{DElia:2016kpz}, which discusses the analogous setup with magnetic fields in small volumes.}. However, such simulations are always naturally performed in the grand canonical ensemble, characterized by a chemical potential and not in the canonical one, which would fix the equilibrium charge distribution, see Sec.~\ref{sec:chargedistr}. This implies that there is a change of relevant ensembles between $E=0$ and any $E\neq 0$, signalled by an abrupt discontinuity in physical observables in the infinite volume limit.

We demonstrate this effect for the grand canonical free energy density $\Omega=-T/V\log\det(\slashed{D}+m)$ in the free case for two different setups. First, homogeneous imaginary electric fields satisfying periodic boundary conditions, where the field is quantized as $eE=2\pi N_E T/L$ with $N_E\in\mathds{Z}$. Second, oscillatory imaginary fields with profile $E(x_1)=E\cos(k x_1)$, again with periodic boundary conditions so that $k=2\pi N_k/L$ and $N_k\in\mathds{Z}$. The determinant is calculated exactly by diagonalizing the Dirac operator on two-dimensional subspaces.
Fig.~\ref{fig:iedep} shows the negative of $\Delta\Omega=\Omega(E)-\Omega(0)$ for both setups on $200^3\times 20$ and $300^3\times 20$ lattices at $T/m=12.5$. In the quantized case, the data at discrete electric field values collapse on a line at $E>0$ that clearly has a different intersect as the $E=0$ value. In the oscillatory case, the curves also approach a discontinuity at zero field as $k$ decreases (either by reducing $N_k$ or by increasing $L$). To eliminate this discontinuity, the equilibrium construction at every nonzero $E$ would have to be performed according to Eq.~\eqref{eq:localLegendre}. The effects of the discontinuity will also be inherited by other observables derived from $\Omega$.

\section{Conclusion}

We calculated the leading order electric response of a non-interacting charged thermal medium using the small-$E$ expansion. Our calculation extends the results of Weldon \cite{Weldon:1982aq} to massive fermions, but more importantly, we take into account another source of $E$-dependence in the free energy of the system. Since the electric field exerts a force on the charges in the medium, an equilibrium can only be reached if there is an equal force of opposite sign to cancel this effect. This can be any force, e.g.\ the strong force in a QCD medium or the electric repulsion in the case of QED. In our case of non-interacting fermions it solely stems from diffusion forces, or in other words the degeneracy pressure. This means that for any electric field profile, a non-trivial charge density distribution develops and keeps the medium in equilibrium. The electric susceptibility has to take this into account, by considering the free energy change added by the redistribution of the charge density. On a more pragmatic level, we find that this contribution is needed to make the susceptibility infrared finite, otherwise one would find a divergence in the homogeneous electric field limit. Our method of calculation can also be applied to magnetic fields, however in this case such contributions do not arise as magnetic fields do not lead to non-trivial equilibrium charge distributions.

Comparing our method to results obtained by using the Schwinger fermion propagator in a homogeneous electromagnetic background, the details of which are described in Ref.~\cite{Elmfors:1998ee}, we find that the
two differ by an ordering of limits. Within the small-$E$ expansion, the electric field approaches zero first, followed by the infinite volume limit (or, infinite wavelength limit for oscillatory fields). In contrast, the Schwinger propagator method involves a direct calculation in the infinite volume for homogeneous fields, and a zero field limit at the end. The two methods give rise to electric susceptibilities that differ by a finite term, see the corresponding high-temperature expansions in Eqs.~\eqref{eq:WeldonxiHTE} and~\eqref{eq:SchwingerxiHTE}.
In turn, since magnetic fields do not involve such an infrared divergence, the two limits can be exchanged for the magnetic susceptibility and, accordingly, both approaches give the same result. 

Our method can be generalized to include QCD interactions. In that case, the diffusion forces
are accompanied by the strong force, affecting the equilibrium charge distribution and, thus, 
leading to corrections in the electric susceptibility. These QCD effects may be taken into account 
via perturbation theory (at high temperatures) or in full lattice QCD simulations.
One advantage of our approach is that it allows for a straightforward implementation on the lattice, similarly to the magnetic field case, by means of current-current correlators~\cite{Bali:2020bcn}.
The analogous density-density correlators for the electric response have also been explored in Ref.~\cite{Endrodi:2021qxz}. In contrast, QCD simulations at nonzero electric fields would not be possible using standard Monte Carlo methods due to the complex action problem. 

Finally, we stress that the calculations presented in this paper (for both approaches) only concern the real part of the thermal free energy, and we do not discuss the imaginary part at $T=0$.
It is important to distinguish the two. The thermal effect is the polarization of charges that are generated due to thermal fluctuations, while the $T=0$ effect is the decay of the vacuum due to virtual charges via the Schwinger mechanism~\cite{Schwinger:1951nm}.

\noindent
\acknowledgments
This research was funded by the DFG (Emmy Noether Programme EN 1064/2-1 and SFB TRR
211 -- project number 315477589).
The authors are grateful to Holger Gies and Bo-Sture Skagerstam for their enlightening insight on calculations of the effective action and 
to Andrei Alexandru, Bastian Brandt, S\'andor Katz, Tam\'as Kov\'acs and Zsolt Sz\'ep for 
helpful discussions.

\appendix
\section{Calculation of the photon polarization tensor\label{app:Pi00}}

We are looking at the real and non-zero temperature part of $\Pi_{00}^{T\neq0}$ first, therefore we need to evaluate the integral
\be
\Re \,\Pi_{00}^{T\neq0} (k_0,k) = -2 e^2\Re\left\{\int \frac{d^4 P}{(2\pi)^4} \Tr \left[\gamma_0(\slashed{P}+\slashed{K}+m)\gamma_0(\slashed{P}+m)\right]\frac{\pi}{E_p}\frac{n_F(|p_0|)\delta(P^2-m^2)}{(P+K)^2+m^2+i\epsilon}\right\}\,,
\ee
where we introduced $E_p = \sqrt{p^2+m^2}$ and the trace is in Dirac space. Evaluating the latter together with the trivial $p_0$ integral allows us to take the $k_0\to0$ limit to find a manifest real expression as the sum of a complex number and its conjugate. Then
\be
\Re \,\Pi_{00}^{T\neq0} (k_0 = 0,k) = 8 e^2\int \frac{d^3\mathbf{p}}{(2\pi)^3} \frac{n_F(E_p)}{E_p}\frac{(2E_p^2+\mathbf {k p})(k^2+2\mathbf {k p})}{(k^2+2\mathbf {k p})^2+4E_p^2\epsilon^2}\,.
\ee
Carrying out the angle integrals and the trivial $\epsilon\to0$ limit yields
\be
\label{eq:Pi_at_k}
\Re \,\Pi_{00}^{T\neq0} (0,k) = -\frac{e^2}{\pi^2}\int_0^\infty dp\,p^2\frac{n_F(E_p)}{E_p}\left[-2+\frac{4E_p^2-k^2}{4kp}\log\frac{(k-2p)^2}{(k+2p)^2}\right]\,.
\ee

In order to carry out the limit in \eqref{eq:xiDefWPi}, $\Re \,\Pi_{00}^{T\neq0} (k_0 = 0,k)$ needs to be expanded around $k=0$ to quadratic order. The leading order can simply be obtained by taking the limit $k\to0$ and yields
\be
\label{eq:Pi00_00}
\Re \,\Pi_{00}^{T\neq0} (0,k = 0) = 2\frac{e^2}{\pi^2}\int_0^\infty dp\,n_F(E_p)\left(\frac{p^2}{E_p}+E_p\right)\,.
\ee
As a check we expand this result around infinite temperature using standard HTE tools and recover the well known result for the photon mass in a plasma (see, e.g., Ref.~\cite{Weldon:1982aq}),
\be
\label{eq:Pi00_00_HTE}
\Re \,\Pi_{00}^{T\neq0} (0,0) = \frac{e^2T^2}{3}-\frac{e^2m^2}{2\pi^2}-\frac{7e^2\zeta'(-2)}{4\pi^2}\frac{m^4}{T^2}+{\cal O}\left(\frac{m^6}{T^4}\right)\,.
\ee

The ${\cal O}(k^2)$ term in the expansion of the polarization tensor needs a more careful analysis. Taking the second derivative with respect to $k$ can be replaced in this case by a $k^2$ derivative and a multiplication by $2$ (the difference is higher order in $k$), leading to
\be
\label{eq:ddkPi_at_k_naiv}
\frac{\partial^2 \Re \,\Pi_{00}^{T\neq0} (0,k)}{\partial k^2} = -\frac{e^2}{\pi^2}{\cal P}\int_0^\infty dp\,\frac{p}{4k^3}\frac{n_F(E_p)}{E_p}\left((4E_p^2+k^2)\log\frac{(k+2p)^2}{(k-2p)^2}-\frac{8kp(k^2-4E_p^2)}{k^2-4p^2}\right)+{\cal O}(k^2)\,,
\ee
where ${\cal P}\int$ denotes principal value integration. Naively taking $k\to0$ leads to an infrared divergent integral, arising from the non-commutativity of the principal value integral and the $k\to0$ limit, since for vanishing external momentum the pole is shifted onto the lower border of the integration. One particular way of handling this problem is by adding
\be
\label{eq:ddkPi_at_k_corr}
0=-\frac{e^2}{\pi^2}{\cal P}\int_0^\infty dp\,\frac{4}{3}\frac{n_F(E_{k/2})}{E_{k/2}}\frac{m^2}{k^2-4p^2}\,.
\ee
Then the $k\to0$ limit of the sum of \eqref{eq:ddkPi_at_k_naiv} and \eqref{eq:ddkPi_at_k_corr} is
\be
\label{eq:ddkPi_IRsafe}
\frac{\partial^2 \Re \,\Pi_{00}^{T\neq0} (0,k)}{\partial k^2}\Bigg|_{k=0} = -\frac{e^2}{\pi^2}\int_0^\infty dp\,\left(\frac{n_F(E_p)}{E_p}-\frac{E_p n_F(E_p)-m n_F(m)}{3p^2}\right)\,,
\ee
which is well behaved around $p=0$. One can do an integration by parts on the term proportional to $1/3p^2$ and change the integration variable to $\omega=E_p$ to cast the final result into the simpler form
\be
\frac{\partial^2 \Re \,\Pi_{00}^{T\neq0} (0,k)}{\partial k^2}\Bigg|_{k=0} = -\frac{e^2}{3\pi^2}\int_m^\infty \frac{d\omega}{\sqrt{\omega^2-m^2}}\left(2n_F(\omega)-\omega\frac{dn_F(\omega)}{d\omega}\right)\,,
\ee

We also need the sum of the three spatial components $\Pi_S^{T\neq0}$, for the magnetic susceptibility. Before carrying out the Dirac trace, it reads
\be
\Re \,\Pi_S^{T\neq0}(k_0,k) = -\frac{e^2}{2} \Re\left\{\int \frac{d^4 P}{(2\pi)^4} \sum_{i=1}^3\Tr \left[\gamma_i(\slashed{P}+\slashed{K}+m)\gamma_i(\slashed{P}+m)\right]\frac{\pi}{E_p}\frac{n_F(|p_0|)\delta(P^2-m^2)}{(P+K)^2+m^2+i\epsilon}\right\}\,.
\ee
Evaluating the trace, the sum for the spatial components as well as the $p_0$ integral yields for $k_0=0$
\be
\Re \,\Pi_S^{T\neq0}(k_0=0,k) = -2e^2 \Re\left\{\int \frac{d^3p}{(2\pi)^3}\frac{n_F(E_p)}{E_p}\frac{3E_p^2-p^2-\mathbf{pk}-3m^2}{-2\mathbf{pk}+i\epsilon}\right\}\,.
\ee
After carrying out the angle integral, the $\epsilon\to0$ limit of the real part reads
\be
\Re \,\Pi_S^{T\neq0}(0,k) = -\frac{e^2}{2\pi^2} \int_0^\infty dp\,p^2\frac{n_F(E_p)}{E_p}\left(2+\frac{k^2+4p^2}{4kp}\log\frac{(k-2p)^2}{(k+2p)^2}\right)\,.
\ee
Then taking the $k\to0$ limit of the second derivative with respect to $k$ is straightforward, and we obtain
\be
\label{eq:ddkPiS}
\frac{\partial^2\Re \,\Pi_S^{T\neq0}(0,k)}{\partial k^2}\Bigg|_{k=0} = \frac{2e^2}{3\pi^2}\int_0^\infty dp\,\frac{n_F(E_p)}{E_p}\,.
\ee

\bibliographystyle{JHEP}
\bibliography{electric_pt}

\end{document}